\documentclass[conference, 10pt]{IEEEtran}
\IEEEoverridecommandlockouts
\usepackage{amsmath}
\usepackage{cite}
\usepackage{amsmath,amssymb,amsfonts}
\usepackage[ruled,linesnumbered,lined]{algorithm2e}
\usepackage{graphicx}
\usepackage{textcomp}
\usepackage{xcolor}
\usepackage{bm}
\usepackage{booktabs}
\usepackage{caption,subcaption}

\usepackage{amsthm,amssymb,amsfonts}

\SetKwInput{Input}{input}
\SetKwInput{Output}{output}
\newtheorem{definition}{Definition}
\usepackage{array}
\newcolumntype{L}[1]{>{\raggedright\let\newline\\\arraybackslash\hspace{0pt}}m{#1}}
\newcolumntype{C}[1]{>{\centering\let\newline\\\arraybackslash\hspace{0pt}}m{#1}}
\newcolumntype{R}[1]{>{\raggedleft\let\newline\\\arraybackslash\hspace{0pt}}m{#1}}

\newtheorem{proposition}{Proposition}

\def\BibTeX{{\rm B\kern-.05em{\sc i\kern-.025em b}\kern-.08em
		T\kern-.1667em\lower.7ex\hbox{E}\kern-.125emX}}

\begin{document}
	\title{A Novel Q-stem Connected Architecture for Beyond-Diagonal Reconfigurable Intelligent Surfaces}
	\author{
		\IEEEauthorblockN{Xiaohua Zhou$^{\ddagger}$,  Tianyu Fang$^{\dagger}$,
			Yijie Mao$^{\ddagger}$ }
		\IEEEauthorblockA{
			$^{\ddagger}$School of Information Science and Technology, ShanghaiTech University, Shanghai 201210, China \\
			$^{\dagger}$Centre for Wireless Communications, University of Oulu, Finland \\
			Email:\{zhouxh3, 
			maoyj\}@shanghaitech.edu.cn, tianyu.fang@oulu.fi
		}
		\thanks{This work has been supported in part by the National Nature Science Foundation of China under Grant 62201347; and in part by Shanghai Sailing Program under Grant 22YF1428400. The work of Tianyu Fang was supported by the Research Council of Finland through 6G Flagship under Grant 346208 and through project DIRECTION under Grant 354901. \textit{(Corresponding author: Yijie Mao)}}
		\\[-2.5 ex]
	}
	\maketitle

	\thispagestyle{empty}
	\pagestyle{empty}
	\begin{abstract}
			Beyond-diagonal reconfigurable intelligent surface (BD-RIS) has garnered significant research interest recently due to its ability to generalize existing reconfigurable intelligent surface (RIS) architectures and provide enhanced performance through flexible inter-connection among RIS elements. However, current BD-RIS designs often face challenges related to high circuit complexity and computational complexity, and there is limited study on the trade-off between system performance and circuit complexity. To address these issues, in this work, we propose a novel BD-RIS architecture named Q-stem connected RIS that integrates the characteristics of existing single connected, tree connected, and fully connected BD-RIS, facilitating an effective trade-off between system performance and circuit complexity. Additionally, we propose two algorithms to design the RIS scattering matrix for a Q-stem connected RIS aided multi-user broadcast channels, namely,  a low-complexity least squares (LS) algorithm and a suboptimal LS-based quasi-Newton algorithm. Simulations show that the proposed architecture is capable of attaining the sum channel gain achieved by fully connected RIS while reducing the circuit complexity. Moreover,  the proposed LS-based quasi-Newton algorithm significantly outperforms the baselines, while the LS algorithm provides comparable performance with a substantial reduction in computational complexity.
  \end{abstract}
	
	\begin{IEEEkeywords}
		Beyond-diagonal reconfigurable intelligent surface (BD-RIS), scattering matrix design.
	\end{IEEEkeywords}
	
	\section{Introduction}
	Reconfigurable intelligent surface (RIS) has emerged as a pivotal technology for advancing 6G networks. By incorporating multiple passive and reconfigurable elements to manipulate signal direction and intensity, RIS significantly enhances received signal quality, thereby improving overall network performance and facilitating the realization of ambitious 6G goals \cite{Wu2020}.
		Recently, beyond-diagonal RIS (BD-RIS) has been proposed in \cite{shen2021modeling} as a pioneering RIS architecture that generalizes conventional diagonal RIS. Pursuant to the microwave theory, BD-RIS introduces a beyond-diagonal scattering matrix (also known as passive beamforming matrix), allowing for interconnections among its elements. 
  
  According to different interconnection architectures, BD-RIS can be categorized into: single,  fully, and group connected architectures \cite{hongyu2023}. 
  Specifically, single connected RIS \cite{wu2019intelligent} refers to the conventional diagonal RIS configuration where each RIS element is connected to a single load or impedance network. In this architecture, each RIS element operates independently, allowing for phase adjustments but limiting the inter-element interactions. Fully connected RIS \cite{shen2021modeling} refers to an architecture in which every RIS element  is interconnected with all other elements. This means that each RIS port can communicate with all other ports, allowing for maximum flexibility in controlling the phase and amplitude of the reflected signals. To reduce the circuit complexity, group connected RIS  is introduced in \cite{shen2021modeling}. It assigns the RIS elements into defined groups where only the elements within each group are interconnected, creating a block diagonal susceptance matrix that allows for independent control and optimization of each group while reducing overall circuit complexity.   Moreover, a tree connected architecture is introduced in \cite{Matteo2024} based on graph theory. By employing a tree graph, tree connected RIS further enhances the trade-off between performance and circuit complexity.
  Besides, simultaneously transmitting and RIS (STAR-RIS) \cite{STARRISxu} and multi-sector BD-RIS \cite{hongyu2023jasc} are  advanced types of RIS that differ from the aforementioned RIS by allowing signals to be both reflected and transmitted through the surface. In this paper, we focus on BD-RIS within reflect mode only.

 Among all the aforementioned BD-RISs, the fully connected BD-RIS enables all elements to connect with each other, thereby achieving the maximum performance gain, albeit at the cost of the highest circuit complexity. Group connected RIS significantly reduces circuit complexity but suffers from performance degradation, failing to achieve an optimal trade-off between channel gain performance and circuit complexity. Although tree connected RIS has been shown to be the simplest BD-RIS architecture for achieving the optimal performance in single-user multiple-input single-output (MISO) \cite{Matteo2024}, its performance in  multi-user MISO remains unexplored.

\par 
Based on these literature, this paper aims to further enhance the trade-off between the channel gain performance and circuit complexity. To this end, we propose a brand-new BD-RIS architecture, named Q-stem connected RIS for multi-user downlink transmission networks. This structure is capable of achieving the performance of fully connected RIS while significantly reducing circuit complexity under controlled complexity conditions. In addition, we propose a closed-form least square (LS)-based algorithm and an LS-based quasi-Newton algorithm to design the scattering matrix, with the aim of maximizing the sum channel gains for our proposed Q-stem connected multi-user MISO.

	
	\section{System  Model and Problem Formulation}\label{Sec:system model}
	
	\subsection{System Model}
	
	Consider a BD-RIS assisted multi-user downlink communication network comprising a base station (BS) with $ L $ transmit antennas, a BD-RIS with $ N $ passive reflecting elements, and a set of $ K $ single-antenna users. The BD-RIS elements are indexed by $ \mathcal{N} = \{1, \cdots, N\} $, while the users are indexed by $ \mathcal{K} = \{1, \cdots, K\} $. The passive beamforming matrix of the BD-RIS is denoted by $ \bm \Theta \in\mathbb C^{N\times N}$. The BS serves all $ K $ users simultaneously with the assistance of this BD-RIS, assuming that the direct links between the BS and users are blocked. The channel between the BD-RIS and user $ k $ is denoted by $ \mathbf h_k\in\mathbb C^{N\times 1} $, and the channel between the BS and the BD-RIS is denoted by $ \mathbf E\in\mathbb C^{N\times L} $. Therefore, the effective channel between the BS and user $k$ is $\mathbf{f}_k^H= \mathbf h_k^H\bm\Theta\mathbf E$. By further defining $ \mathbf H\triangleq [\mathbf h_1,\cdots,\mathbf h_K]\in\mathbb C^{N\times K} $,  the effective channels for all users can be written in a compact form as $\mathbf H^H\mathbf \Theta\mathbf E$. In this work, our goal is to design the passive beamforming matrix of a novel BD-RIS architecture named Q-stem connected BD-RIS, so as to maximize the sum of effective channel gains among users, i.e., $\sum_{k=1}^K\|\mathbf h_k^H\mathbf \Theta\mathbf E\|^2$. It is equivalent to maximize $\|\mathbf H^H\mathbf \Theta\mathbf E\|_F^2$. 
	
\textit{BD-RIS Modeling}:
According to \cite{shen2021modeling}, an $N$-element BD-RIS can be modeled as $N$ antennas connected to an $N$-port reconfigurable impedance network, equipped with tunable passive impedance components. 
Following microwave network theory \cite{pozar2021microwave}, the 
$N$-port reconfigurable impedance network can be characterized by Y-parameters based on a symmetric admittance matrix that satisfies $ \mathbf{Y}=j \mathbf{B} $, where the susceptance matrix $\mathbf{B} \in \mathbb{R}^{N \times N} $ satisfies $ \mathbf{B}=\mathbf{B}^T$.
The scattering matrix $ \bm\Theta $ is therefore represented by
\cite{Universal2024}: 
	\begin{equation}\label{eq:theta}
		\bm \Theta=\left(\mathbf{I}_N+jZ_0 \mathbf{B}\right)^{-1}\left(\mathbf{I}_N-jZ_0 \mathbf{B}\right),
	\end{equation}
	where $ Z_0 $ is the reference impedance used for computing the scattering parameter.
    The modeling in (\ref{eq:theta}) ensures the scattering matrix $\bm \Theta$ to be symmetric unitary,  following a lossless and reciprocal circuit network. 
    Depending on the circuit topology of the 
$N$-port reconfigurable impedance network, the scattering matrix $\bm \Theta$—particularly the matrix  $\mathbf{B}$ in $\bm \Theta$—is required to satisfy different constraints, resulting in different BD-RIS architectures.

	\subsection{Proposed Q-Stem Connected RIS}
 In this subsection, we propose a novel BD-RIS architecture that combines the characteristics of single connected, tree connected, and fully connected RIS structures.
 Following \cite{Matteo2024}, we employ graph theoretical tools to model the general circuit topology of our proposed Q-stem connected RIS.

\begin{definition}{(Q-stem connected RIS)}\label{def:Q-stem}
We use a graph to denote the circuit topology of Q-stem connected RIS, which is given as:
	\begin{equation}\label{eq:graph-Q}
		\mathcal{G}=(\mathcal{V},\mathcal{E}),
	\end{equation}
	where $ \mathcal{V} $ is the vertex set of $ \mathcal{G} $ and it corresponds to the index set of RIS ports, i.e.,
	\begin{equation}\label{key}
		\mathcal{V} = \left\{1, 2, \cdots, N \right\},
	\end{equation}
	and $ \mathcal{E} $ is the edge set of $ \mathcal{G} $ indicating the connection between ports. 
Furthermore, we define the set of the first $Q$ ($Q\leq N$) vertices as $\mathcal{V}_Q$, i.e., $\mathcal{V}_Q=\{1,2,\cdots,Q\}$, and the remaining vertices as $\mathcal{V}_{\Bar{Q}}$, i.e. $\mathcal{V}_{\Bar{Q}}=\{Q+1,\cdots,N\}$.  

$Q$-stem connected RIS uses a tunable admittance to connect the $n$-th port and the $m$-th port when $n \in \mathcal{V}_Q, m \in \mathcal{V}$ or $n \in \mathcal{V}_{\Bar{Q}}, m \in \mathcal{V}_Q$. Therefore, its edge set is given as:
	\begin{equation}\label{eq:edge-Q}
		\begin{split}
		    \mathcal{E} = \left\{(n,m)|n \in \mathcal{V}_Q, m \in \mathcal{V} \, \text{or} \, \ n \in \mathcal{V}_{\Bar{Q}}, m \in \mathcal{V}_Q, \right.  \\
      \left. [\mathbf{B}]_{n,m} \neq 0, n \neq m  \right\}.
		\end{split}
	\end{equation}
Here, $[\mathbf{B}]_{n,m} \neq 0$ implies there is an edge between vertex $ n $ and vertex $ m $ if and only if there is a tunable admittance connecting the $n$-th port and the $m$-th port. 	
\end{definition}
Let $\mathcal{B}_Q$ represent the set of all feasible susceptance matrices $\mathbf{B}$ for Q-stem connected RIS. Equation \eqref{eq:edge-Q} can then be equivalently reformulated as a constraint for $\mathbf{B}$ as follows:  	
\begin{equation}\label{eq:B-arrow}
		\mathcal{B}_Q=\left\{ \mathbf{B}| 	[\mathbf{B}]_{n,m}=0, n \neq m, n > Q, m > Q \right\}.
	\end{equation}

		 \begin{figure}[t!]
	\centering
	 		\begin{subfigure}{1\linewidth}
	 			\centering
     \includegraphics[width=0.6\linewidth]{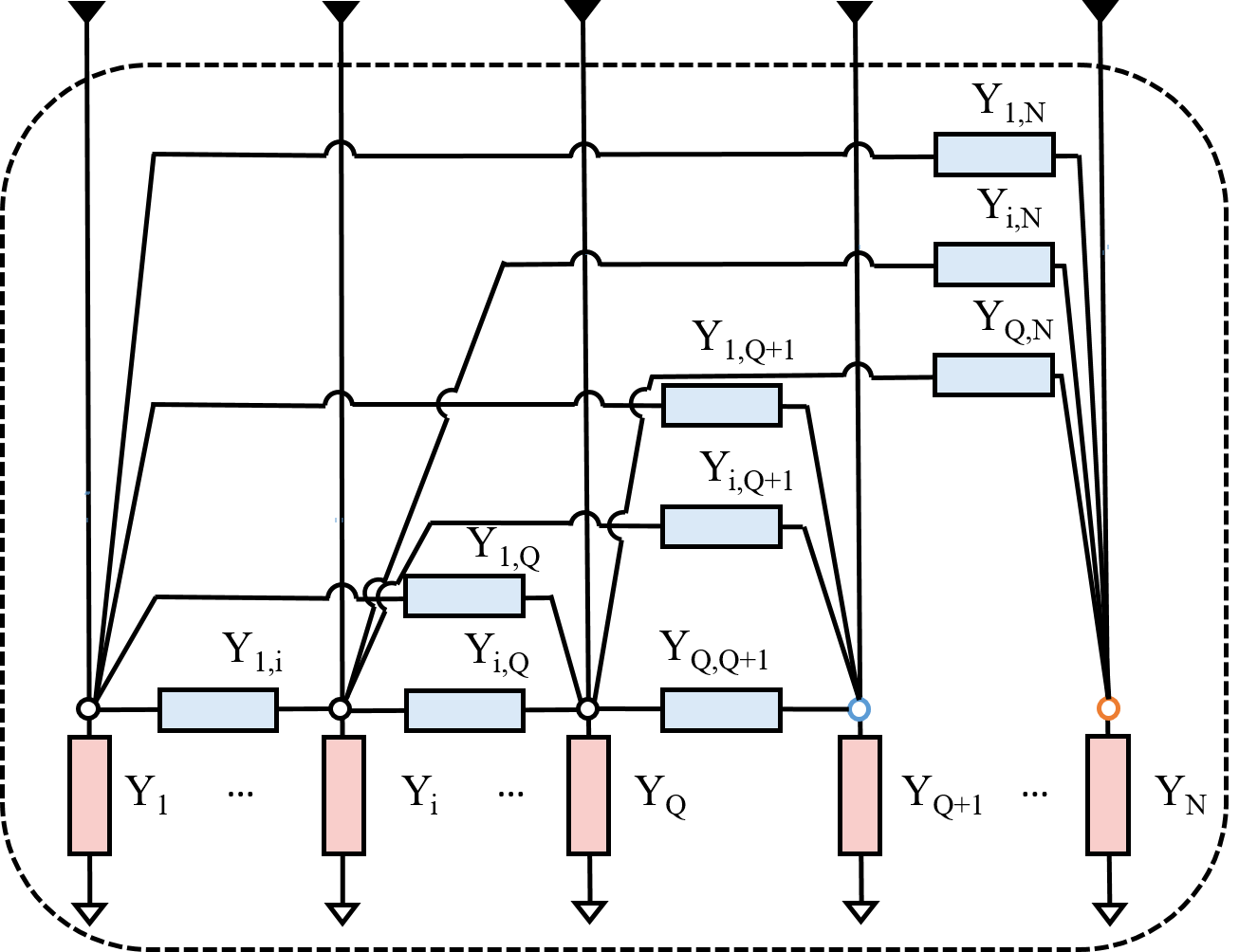}
	 			\caption{The architecture of Q-stem connected RIS.}
	 			\label{fig:port}
	 		\end{subfigure}
    \vspace{1mm}

	 		\begin{subfigure}{1\linewidth}
    \centering 
	 			\includegraphics[width=0.6\linewidth]{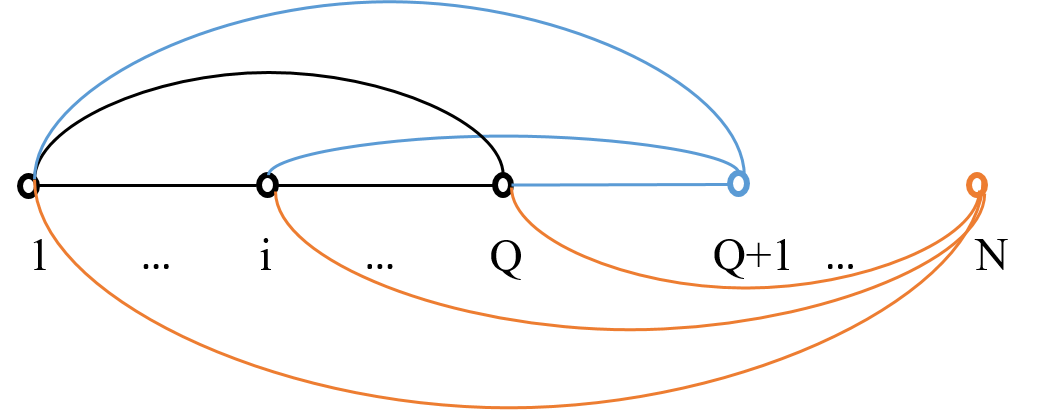}
	 			\caption{Graph representation for Q-stem connected RIS.}
	 			\label{fig:graph}
	 		\end{subfigure}
	 	\caption{The circuit architecture of Q-stem connected RIS and its graph representation.}
	 	\label{fig:Q-stem-architecture}
	 \end{figure} 
	
\begin{figure}
		\centering
		\includegraphics[width=0.46\linewidth]{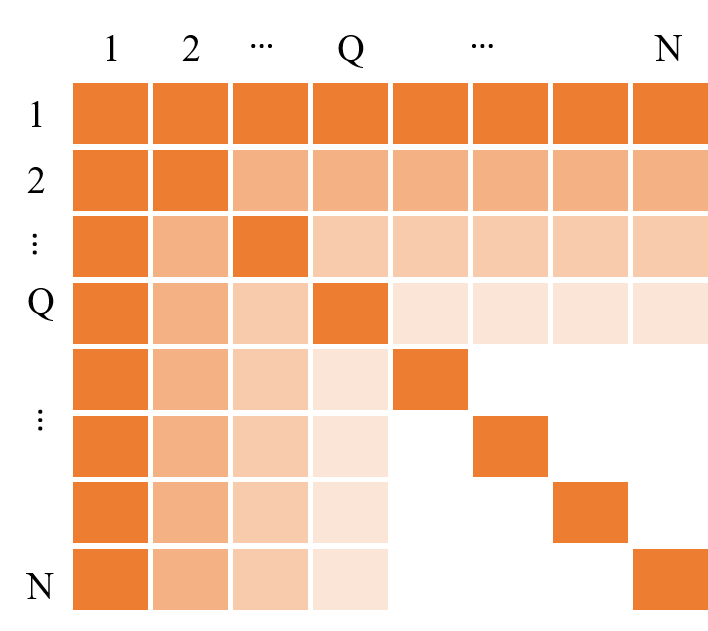}
		\caption{The shape of a feasible
susceptance matrix $ \mathbf{B} $ for Q-stem connected RIS. 
The non-zero elements in $\mathbf{B}$ are represented in orange.}
		\label{fig:arrow}
  \vspace{-4mm}
	\end{figure}

Fig. \ref{fig:Q-stem-architecture} further illustrates the circuit architecture and the graph representation for our proposed Q-stem connected RIS in Definition \ref{def:Q-stem}. Specifically, in the circuit architecture shown in Fig. \ref{fig:port}, the first $Q$ ports connect with all other ports while the remaining ports only connect to the first $Q$ ports.
From a graph theory perspective, the degree of vertices in $\mathcal{V}_Q$ is $N-1$, while the degree of the vertices in $\mathcal{V}_{\Bar{Q}}$ is $ Q $, as depicted in Fig. \ref{fig:graph}. 
In Fig. \ref{fig:arrow}, we further illustrate a feasible susceptance matrice $\mathbf{B}$ for Q-stem connected RIS which satisfies  the constraint in \eqref{eq:B-arrow}, i.e., $\mathbf{B} \in \mathcal{B}_Q$. All
non-zero elements in $\mathbf{B}$ are represented in orange color.
Obviously, $\mathbf{B}$ has in total $ (2N-1)Q-Q^2 $ non-zero off-diagonal entries.

\addtolength{\topmargin}{0.03in}	

	\subsection{Connection Between Q-stem Connected RIS and Other  RIS Architectures}
	Q-stem connected RIS is a general BD-RIS architecture which bridges existing single connected RIS, tree connected RIS, and fully connected RIS by adjusting the parameter $Q$. Specifically,
	\begin{itemize}
		\item When $ Q=0 $, the proposed Q-stem connected RIS reduces to a single connected RIS, the corresponding graph is empty. The susceptance matrix $\mathbf{B}$ reduces to: 
  \begin{equation}
      \mathbf{B}=\mathrm{diag}([\mathbf B]_{11},\cdots,[\mathbf B]_{NN}).
  \end{equation}
		\item When $ Q=1 $, the proposed Q-stem connected RIS is equivalent to a tree connected RIS. Its graph becomes a tree. The susceptance matrix $\mathbf{B}$ for tree connected RIS becomes
  \begin{equation}
          \mathbf{B}=\mathbf{B}^T,
			\mathbf{B} \in \mathcal{B}_G,
  \end{equation}
   where 
	$\mathcal{B}_G=\left\{ \mathbf{B}| 	[\mathbf{B}]_{n,m}=0, n \neq m, m > 1, n > 1  \right\}$.
		\item When $ Q=N-1 $, the proposed Q-stem connected RIS becomes a fully connected RIS, where the graph is complete. In this case, all ports are interconnected via tunable impedance components, offering maximum design flexibility and performance \cite{shen2021modeling}.
	\end{itemize}

\subsection{Circuit Complexity of Q-Stem Connected RIS}
The circuit topology complexity is reflected in the number of independent tunable admittance components, which corresponds to the number of independent variables in the symmetric susceptance matrix $\mathbf{B}$. Q-stem connnected RIS includes $ N $ admittance components connecting each port to ground and $ QN-{Q(Q+1)}/{2} $ admittance components interconnecting the ports to each other, yielding a total of $ QN+N-{Q(Q+1)}/{2} $ admittance components. Table \ref{tab:circuit-complexity} provides the circuit complexity comparison among various architectures for BD-RIS. Notably, the circuit complexity of Q-stem connected RIS scales linearly with the number of elements for a fixed $Q$.
	
	\begin{table}[t!]
	\begin{center}
		\caption{Circuit Complexity Comparison for Different RIS Architectures.}
		\label{tab:circuit-complexity}
		\begin{tabular}{ccc}
			\toprule
			\textbf{Architecture}& 	\textbf{Reference} & \textbf{Circuit Complexity}  \\
			\hline\hline
			Single connected RIS & \cite{wu2019intelligent}&$ N $  \\
			\hline
			Group connected RIS & \cite{shen2021modeling}&${N}(N/G+1)/2 $  \\
			\hline
			Fully connected RIS & \cite{shen2021modeling} &${N(N+1)}/2$\\
			\hline
			Tree connected RIS & \cite{Matteo2024}&$ 2N-1 $\\
			\hline
			Q-Stem connected RIS &Proposed& $ QN+N-{Q(Q+1)}/2 $\\
			\bottomrule
		\end{tabular}
	\end{center}
 \vspace{-4mm}
\end{table}

\subsection{Problem Formulation}
To show the effectiveness of our proposed Q-stem connected RIS, we next design its scattering matrix $\bm \Theta$ with the aim of maximizing the sum channel gains among users. The formulated problem is given as:
	\begin{subequations}\label{eq:pro-formula}
		\begin{align}
			\max_{\bm \Theta} \quad &\|\mathbf H^H\mathbf \Theta\mathbf E\|_F^2 \label{subeq:obj}\\
			\operatorname{s.t.} \quad &\bm \Theta=\left(\mathbf{I}+jZ_0\mathbf{B}\right)^{-1}\left(\mathbf{I}-jZ_0\mathbf{B}\right), \label{eq:theta-con}\\
			&\mathbf{B}=\mathbf{B}^T,\label{eq:B-con}\\
			&\mathbf{B} \in \mathcal{B}_Q,\label{eq:B-stru}
		\end{align}
	\end{subequations}
 where $\|\cdot\|_F$ denotes the Frobenius Norm. Problem \eqref{eq:pro-formula} is challenging to solve since constraint \eqref{eq:theta-con} is highly non-convex. 
 Although the passive beamforming design to maximize the channel gain has been developed for fully connected RIS \cite{Santamaria2023,fang2023low}, tree connected RIS \cite{Matteo2024} and group connected RIS \cite{nerini2023closed}, they cannot be applied to address the problem for our proposed Q-stem connected RIS due to its unique scattering matrix constraint. 
 Specifically, the algorithm in \cite{Santamaria2023,nerini2023closed, Matteo2024} focus in single-user transmission networks while \cite{fang2023low} only address fully connected RIS configurations.
To the best of our knowledge, no existing algorithms proposed for BD-RIS can be directly applied to solve it. In the next section, we will propose an efficient algorithm to address this problem.

	\section{Proposed Efficient Algorithm}

 In this section, we first analyze an upper bound of the sum channel gains for problem \eqref{eq:pro-formula}.
 Based on such upper bound, we then propose a closed-form solution to problem \eqref{eq:pro-formula}. 
	\subsection{Theoretical Upper Bound Analysis for Problem \eqref{eq:pro-formula}}
  \label{sec:Upperbound}
 In this subsection, we first relax problem \eqref{eq:pro-formula} by removing constraints \eqref{eq:B-con} and \eqref{eq:B-stru}. 
 Then, we apply singular value decomposition (SVD) to $ \mathbf{H}^H $ and $ \mathbf{E} $, i.e., $ \mathbf{H}^H= \mathbf{U}\mathbf{S}\mathbf{V}^H$ and $ \mathbf{E} = \mathbf{P}\bm \Sigma\mathbf{W}^H$, where $\mathbf{S} \in \mathbb{R}^{K \times N}$, $\bm \Sigma \in \mathbb{R}^{N \times L}$, and $\mathbf{U} \in \mathbb{C}^{K \times K}$, $\mathbf{V} \in \mathbb{C}^{N \times N}$, $\mathbf{P} \in \mathbb{C}^{N \times N}$,  $\mathbf{W} \in \mathbb{C}^{L \times L}$ are unitary matrices.
 The relaxed problem is obtained as:
	\begin{subequations}\label{eq:SVD}
		\begin{align}
			\max_{\bm \Theta} \quad &\|\mathbf{U}\mathbf{S}\mathbf{V}^H\mathbf \Theta \mathbf{P}\bm \Sigma\mathbf{W}^H\|_F^2\\
			\operatorname{s.t.} \quad \label{con:unitary}&\bm\Theta\bm\Theta^H=\mathbf I_N.
		\end{align}
	\end{subequations}
Problem \eqref{eq:SVD} is also a relaxed formulation of the sum channel gain maximization problem for fully connected RIS, where  the symmetric contraint $\bm\Theta = \bm\Theta^T$ is relaxed.
	Since $ \mathbf{U}\mathbf{U}^H =\mathbf{I}_K $ and $ \mathbf{W}\mathbf{W}^H =\mathbf{I}_L $, the objective function of problem \eqref{eq:SVD} is equivalent to $ \|\mathbf{S}\mathbf{V}^H\mathbf \Theta \mathbf{P}\bm \Sigma\|_F^2 $. To obtain more insights of the channel shapping capability of BD-RIS, we introduce the following definition.
	\begin{definition}(Degree of freedom)\label{definition:DoF}
		The degree of freedom (DoF), also known as multiplexing gain, is the maximum number of independent streams transmitted in parallel over a MIMO channel. The DoF for the compact effective channels $ \mathbf F\triangleq[ \mathbf{f}_1,\cdots, \mathbf{f}_K]\in\mathbb C^{L\times K}$ is defined as  \cite{zhao2024channelshapingusingdiagonal}
		\begin{equation}\label{eq:DoF}
			M=\lim_{\rho \rightarrow \infty} \frac{\log\det\left(\mathbf{I}_L+\rho \mathbf{F}\mathbf{F}^H\right)}{\log \rho}=\min(K,L,N),
		\end{equation}
		where $ \rho $ is the signal-to-noise ratio (SNR).
	\end{definition}
	Based on the definition of DoF $M$ in Definition \ref{definition:DoF}, we then partition the matrices $ \mathbf{V} $ and $ \mathbf{P} $ as $ \mathbf{V} = [\mathbf{V}_{M},\mathbf{V}_{N-M}] $ and $ \mathbf{P} = [\mathbf{P}_{M},\mathbf{P}_{N-M}] $. We also partition $ \mathbf{S} $ and $ \bm \Sigma $ as $\mathbf{S}=\operatorname{ diag}(\mathbf{S}_{M}, \mathbf{S}_{K-M}) $ and $ \bm \Sigma = \operatorname{diag}(\bm \Sigma_M, \mathbf{\Sigma}_{L-M}) $. In this way, we have
	\begin{equation}\label{eq:upper-bound}
		\begin{split}
			\|\mathbf{S}\mathbf{V}^H\bm \Theta \mathbf{P}\bm \Sigma\|_F^2 
			=& \left\| \mathbf S_M\mathbf V_M^H\mathbf\Theta\mathbf P_M\mathbf \Sigma_M \right\|^2_F    \\
			 \overset{\underset{\mathrm{(a)}}{}}{\leq}& \left\|\mathbf{S}_{M}\bm \Sigma_M\right\|_F^2,
		\end{split}
	\end{equation}
	where the equality in $\mathrm{(a)}$ holds if and only if
\begin{equation}\label{eq:optimal-T}
\mathbf{V}_{M}^H\mathbf \Theta \mathbf{P}_{M} = \bm\Phi,
	\end{equation}
 and $\bm\Phi=\mathrm{diag}([e^{j\phi_1},\cdots,e^{j\phi_M}])$ with $\phi_m\in[0,2\pi), m=1,\cdots,M$. This is a sufficient condition to achieve the theoretic upper bound of problem \eqref{eq:pro-formula}. 
 This upper bound can provide more insights in channel rearrangement and channel space alignment, and more details are provided in \cite{zhao2024channelshapingusingdiagonal}. However, it is worth noting that, limited by the physical reciprocal property, i.e., $\mathbf \Theta=\mathbf \Theta^T$, all BD-RIS structures can not attain this theoretical upper bound in most cases as shown in the following Proposition \ref{pro:fully-con}.

\begin{proposition}\label{pro:fully-con}
		Fully connected RIS can not achieve the performance upper bound in \eqref{eq:upper-bound} when the DoF for the compact effective channels $ \mathbf F$ is larger than 1, i.e., $M>1$. It only achieves the performance upper bound when $ M=1 $. 
	\end{proposition}
	\textit{Proof.} 
 See Appendix.
	$\hfill\blacksquare$
 
Since fully connected RIS offers the highest design flexibility among BD-RIS structures and cannot achieve the upper bound in \eqref{eq:upper-bound} when $M>1$, neither can any other BD-RIS structures, including Q-stem connected RIS.

\subsection{Proposed Least Square-based Algorithms for Problem \eqref{eq:pro-formula}}
	

Through the analysis in the last subsection, we obtain that the upper bound in \eqref{eq:upper-bound} for problem \eqref{eq:pro-formula}  cannot be achieved by Q-stem connected RIS. 
As the upper bound \eqref{eq:upper-bound} is guaranteed to be achieved by the sufficient condition \eqref{eq:optimal-T}, we therefore obtain an upper bound of problem \eqref{eq:pro-formula}  by replacing its objective function \eqref{subeq:obj} with \eqref{eq:optimal-T}, which is given as
\begin{equation}\label{eq:upper_bound_of_8}
       \eqref{eq:optimal-T},\eqref{eq:theta-con}, \eqref{eq:B-con}, \eqref{eq:B-stru}.  
\end{equation}
Following our discussion in Section \ref{sec:Upperbound}, by relaxing constraint \eqref{eq:B-stru} in \eqref{eq:upper_bound_of_8}, we obtain problem \eqref{eq:theta-opt} in the appendix for fully-connected RIS. Such problem has no solution when $ M>1 $. Therefore, \eqref{eq:upper_bound_of_8} has no solution when $ M>1 $ either due to the additional constraint \eqref{eq:B-stru}.
In the following, we propose a least square (LS) method for solving \eqref{eq:upper_bound_of_8}.

Before delving into the LS solution, we first introduce a novel vectorization operator. 
	\begin{definition}(Independent Vectorization)\label{definition:vec}
 For any matrix $\mathbf{B}$, the operator 
$\operatorname{vec_i}(\mathbf{B})$ is used to extract all independent variables from $\mathbf{B}$ and convert them into a vector. 
	\end{definition}

 \par For our proposed Q-stem connected RIS, 
 there are $ QN+N-{Q(Q+1)}/{2} $ independent variables in $ \mathbf{B} $  as per Fig. \ref{fig:arrow}. Therefore, the independent vectorization of $ \mathbf{B} $ is given as
\begin{equation} 
		\begin{split}
			\operatorname{vec_i}(\mathbf{B}) &= \left[[\mathbf{B}]_{1,1}, \ldots, [\mathbf{B}]_{1,N}, \ldots, [\mathbf{B}]_{Q,Q}, \ldots, [\mathbf{B}]_{Q,N},\right.\\ &\left.[\mathbf{B}]_{Q+1,Q+1}, \ldots, [\mathbf{B}]_{N,N}\right]^T\in \mathbb{R}^{\left( QN+N-\frac{Q(Q+1)}{2} \right) \times 1}.
		\end{split}
	\end{equation}
 It is worth noting that the independent vectorization $\operatorname{vec_i}(\mathbf{B})$ is a linear transformation of the conventional vectorization operator $\operatorname{vec}(\mathbf{B})$ which simply stacks the columns of  $\mathbf{B}$ on top of one another. 
Specifically, by defining $ \mathbf{R}\in \mathbb{R}^{N^2 \times ( QN+N-\frac{Q(Q+1)}{2} )}$ as a transformation matrix and $\mathbf{b}\triangleq\operatorname{vec_i}(\mathbf{B})$, we have 
\begin{equation}\label{eq:nullspace}
		\operatorname{vec}(\mathbf{B})=\mathbf{R}\mathbf{b},
	\end{equation}
where $\mathbf{R}$ is used to map  $\mathbf{b}$ to  $\operatorname{vec}(\mathbf{B})$ so as to satisfy the  Q-stem connected RIS constraints $\mathbf{B} \in \mathcal{B}_Q$ and $\mathbf{B}=\mathbf{B}^T$. The elements of $\mathbf{R}$ are either 0 or 1. 

Next, we use $\mathbf{b}\triangleq\operatorname{vec_i}(\mathbf{B})$ to obtain our proposed LS method.
Specifically, we first substitute \eqref{eq:theta-con} into \eqref{eq:optimal-T} and equivalently rewrite \eqref{eq:optimal-T} as
	\begin{equation}\label{eq:linear-eq-matrix}
		\mathbf{B} \mathbf{C} = \mathbf{D},
	\end{equation}
	where $ \mathbf{C} =jZ_0 (\mathbf{V}_{M}+ \mathbf{P}_{M}) \in \mathbb C^{N \times M}$ and $ \mathbf{D} = \mathbf{P}_{M}-\mathbf{V}_{M} \in \mathbb C^{N \times M}$.
	Since $\mathbf{B}$ is a real-value matrix, we could further  transform \eqref{eq:linear-eq-matrix} into real and imaginary equations, which is given as 
	\begin{equation}\label{eq:re-im}
			\mathbf{B} \Re\{\mathbf{C}\} = \Re \{\mathbf{D}\},\,\,\,\,
			\mathbf{B} \Im\{\mathbf{C}\} = \Im \{\mathbf{D}\},
	\end{equation} 
where $\Re(\cdot)$ and $\Im(\cdot)$ respectively denote the real and imaginary parts of a complex matrix.

	By applying the conventional vectorization operator to both sides of 
 \eqref{eq:re-im}, \eqref{eq:re-im} can be equivalently transformed into
	\begin{subequations}\label{eq:re-im-vec}
		\begin{align}
			\left(\Re\{\mathbf{C}\}^T  \otimes \mathbf{I}_N\right) \operatorname{vec}(\mathbf{B})  =& \operatorname{vec}(\Re \{\mathbf{D}\}),\\
			\left(\Im\{\mathbf{C}\}^T  \otimes \mathbf{I}_N\right) \operatorname{vec}(\mathbf{B})  =& \operatorname{vec}(\Im \{\mathbf{D}\}),
		\end{align}
	\end{subequations} 
 where $\otimes$ represents the Kronecker product.
We further substitute \eqref{eq:nullspace} into \eqref{eq:re-im-vec} and obtain	\begin{subequations}\label{eq:re-im-b}
		\begin{align}
			\left(\Re\{\mathbf{C}\}^T  \otimes \mathbf{I}_N\right) \mathbf{R}\mathbf{b}  =& \operatorname{vec}(\Re \{\mathbf{D}\}),\label{eq:vecDre}\\
			\left(\Im\{\mathbf{C}\}^T  \otimes \mathbf{I}_N\right) \mathbf{R}\mathbf{b}  =& \operatorname{vec}(\Im \{\mathbf{D}\})\label{eq:vecDim}.
		\end{align}
	\end{subequations} 
	By combining \eqref{eq:vecDre} and \eqref{eq:vecDim}, we equivalently transform the problem \eqref{eq:upper_bound_of_8} into a linear  equation of $\mathbf{b}$ given as:
	\begin{equation}\label{eq:Abz}
		\mathbf{A}\mathbf{b}=\mathbf{z},
	\end{equation}
	where $ \mathbf{A}= \left[\left(\Re\{\mathbf{C}\}^T  \otimes \mathbf{I}\right) \mathbf{R};\left(\Im\{\mathbf{C}\}^T  \otimes \mathbf{I}\right) \mathbf{R} \right] \in \mathbb{R}^{2MN \times \left(\frac{2N-1}{2}Q-\frac{Q^2}{2}+N\right)}$, $ \mathbf{z}=\left[\operatorname{vec}(\Re \{\mathbf{D}\});\operatorname{vec}(\Im \{\mathbf{D}\})\right] \in \mathbb{R}^{2MN\times 1} $.

Due to the equivalence between \eqref{eq:upper_bound_of_8} and \eqref{eq:Abz}, equation \eqref{eq:Abz} has no solution when $ M>1 $. Therefore, we propose to employ the LS solution to solve equation \eqref{eq:Abz} as closely as possible. Following the fundamentals of LS solution in \cite{bjorck1990least}, we obtain the LS solution for \eqref{eq:Abz}, which is given as:	\begin{equation}\label{eq:approximate-solution}
		\mathbf{b} = \left(\mathbf{A}^T\mathbf{A}\right)^{-1}\mathbf{A}^T\mathbf{z}.
	\end{equation}
The details of our proposed LS method is illustrated in Algorithm \ref{alg:low-complexity}.
The computational complexity of Algorithm \ref{alg:low-complexity} is dominated by Step 6, which has the order of $ \mathcal{O}(Q^3N^3) $.

It is worth noting that our proposed LS method is an efficient closed-form solution. As such, it can be effectively utilized as a good initialization point for the quasi-Newton method \cite{shen2021modeling,fang2022fully}, which is a typical approach to solve unconstrained optimization problems. In our proposed LS-based quasi-Newton algorithm, the scattering matrix is first initialized by Algorithm \ref{alg:low-complexity} with $\operatorname{vec}(\mathbf{B}) =\mathbf{R}\mathbf{b}$. This is then followed by applying the quasi-Newton method to address the following problem: 
	\begin{equation}\label{eq:pro-quasi}
			\max_{\mathbf{b}} \quad \|\mathbf H^H \left(\mathbf{I}+jZ_0\mathbf{B}\right)^{-1}\left(\mathbf{I}-jZ_0\mathbf{B}\right)\mathbf E\|_F^2 
	\end{equation}

	\setlength{\textfloatsep}{7pt}	
	\begin{algorithm}[t!]
		\caption{Proposed least spare (LS) solution for problem \eqref{eq:pro-formula}}
		\label{alg:low-complexity}
		\textbf{Input:} The channel matrices $ \mathbf H^H $ and $ \mathbf E $\;
		{
			Obtain $ \mathbf{V} $ and $ \mathbf{P} $ through SVD decompositon;\\
			Obtain $ \mathbf{V}_{M} $ and $ \mathbf{P}_{M} $ defined in \eqref{eq:upper-bound};\\
			Obtain $ \mathbf{R} $, $ \mathbf{C} $ and $ \mathbf{D} $ defined in \eqref{eq:nullspace} and \eqref{eq:linear-eq-matrix};\\
			Obtain $ \mathbf{A} $ and $ \mathbf{z} $ defined in \eqref{eq:Abz};\\
			Obtain $ \mathbf{b} = \left(\mathbf{A}^T\mathbf{A}\right)^{-1}\mathbf{A}^T\mathbf{z} $;\\
		}
	\end{algorithm}

	
	\section{Simulation Results}

	\begin{figure*}
		\begin{center}
			\begin{minipage}{0.32\textwidth}
				\includegraphics[width=1\linewidth]{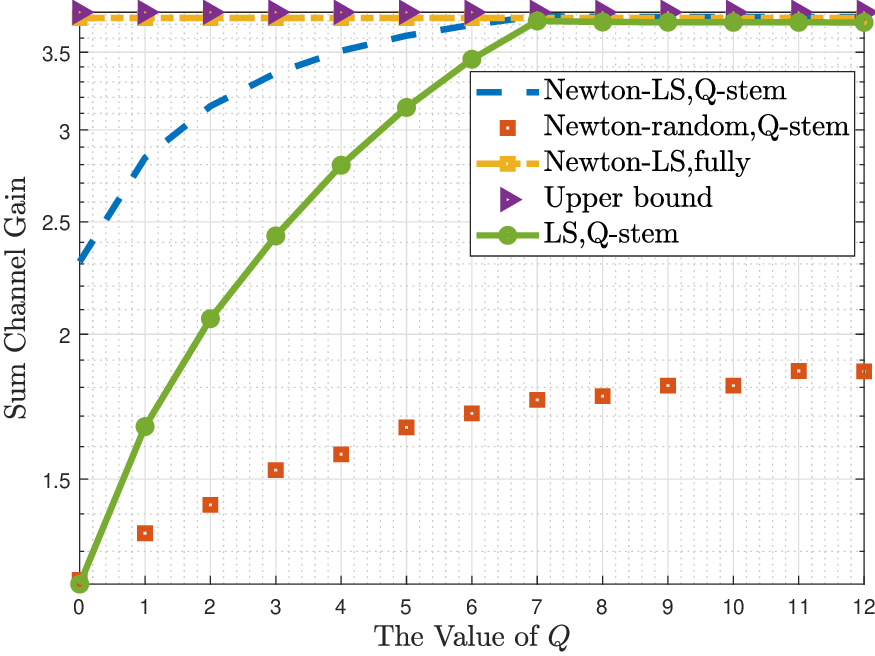}
		\caption{Channel gain versus the value of $Q$ in $\mathbf{B}$ when $L=K=4$ and $N=64$.}
		\label{fig:num_Q}
			\end{minipage}
   \hspace{1mm}
			\begin{minipage}{0.32\textwidth}
				\includegraphics[width=1\linewidth]{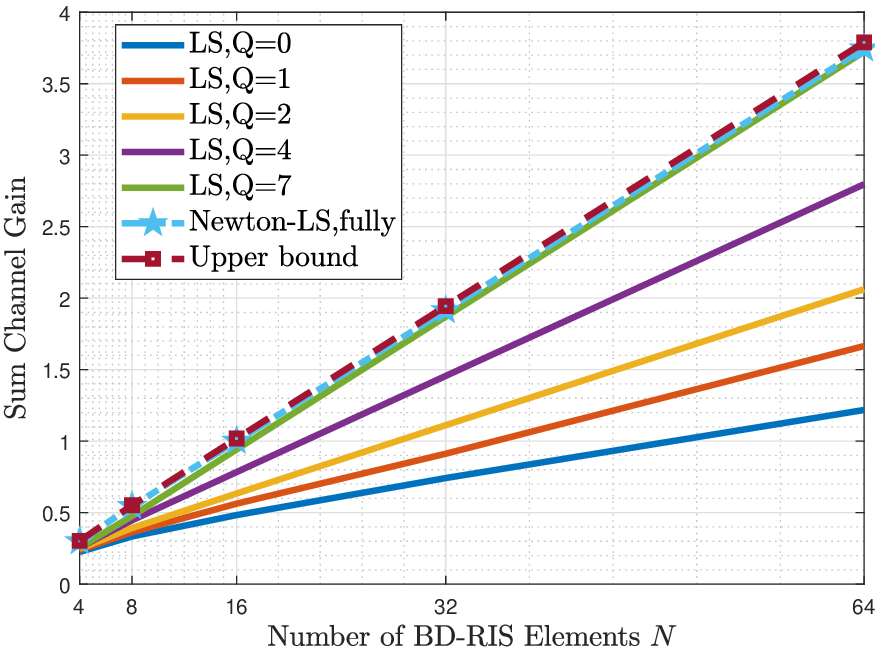}
		\caption{Channel gain versus the number of BD-RIS elements when $L=K=4$.}
		\label{fig:num_RIS}
			\end{minipage}
   \hspace{1mm}
			\begin{minipage}{0.32\textwidth}
				\includegraphics[width=1\linewidth]{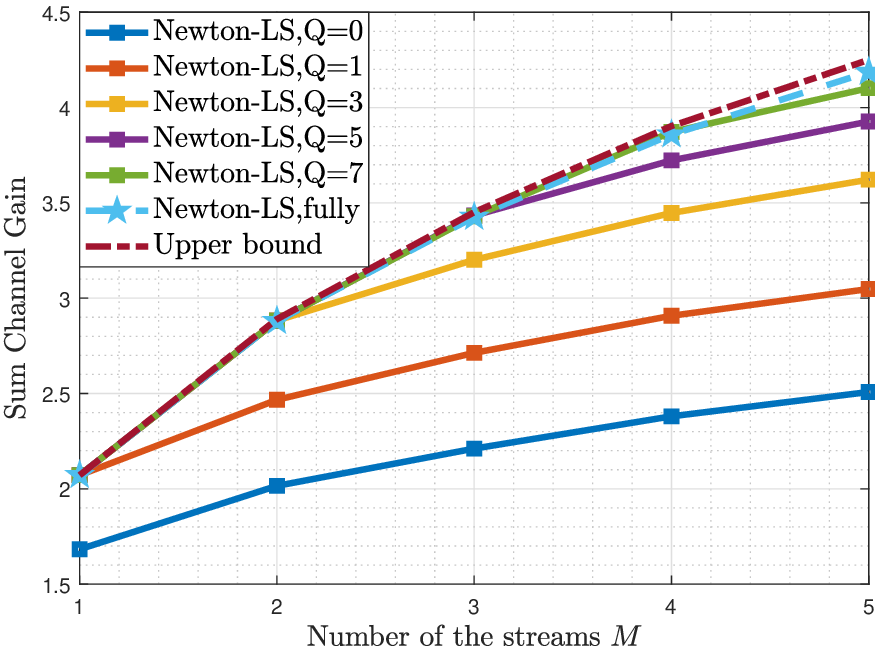}
		\caption{Channel gain versus the number of streams when $L=5$ and $N=64$.}
		\label{fig:L5K}
			\end{minipage}
		\end{center}
  \vspace*{-0.5cm}
	\end{figure*}

In this section, we evaluate the simulation results of the proposed Q-stem connected RIS architecture and the LS algorithm for solving the sum channel gain maximization problem. We assume that the distances between the BS and the RIS, and between the RIS and users are both $50\sqrt{2}$ meters. The path loss follows the model $ P(d)=L_0 d^{-\alpha} $, where $ L_0 = -30 $ dB represents the reference path loss at $ d=1 $ m, $ d $ is the link distance, and $ \alpha $ is the path loss exponent. Following \cite{fang2022fully}, the path loss exponents for the BS-RIS and RIS-user links are set to 2 and 2.2, respectively. The small-scale fading model is based on Rayleigh fading. All simulation results are averaged over 100 random channel realizations. The following five schemes are compared in this work:
\begin{itemize}
    \item \textit{Newton-LS, Q-stem:} This refers to the quasi-Newton method initialized with our proposed LS method for Q-stem connected RIS.
    \item \textit{Newton-random, Q-stem:} This refers to the quasi-Newton method initialized with a random scattering matrix for Q-stem connected RIS.
    \item \textit{Newton-LS, fully:} This refers to the quasi-Newton method initialized with our proposed LS method for fully connected RIS, i.e., $Q=N-1$.
    \item \textit{Upper bound:} This refers to the upper bound $\left\|\mathbf{S}_{M}\bm \Sigma_M\right\|_F$ in \eqref{eq:upper-bound}.
     \item \textit{LS, Q-stem:} This refers to our proposed LS method in Algorithm \ref{alg:low-complexity} for Q-stem connected RIS.
\end{itemize}
For the proposed Q-stem connected RIS, it simplifies to a single connected RIS when $Q=0$, and to tree connected RIS when $Q=1$.




In Fig.~\ref{fig:num_Q}, we show the channel gain versus the value of $Q$ in susceptance matrix $\mathbf{B}$ with $N=64$. While the fully connected RIS is close to achieving upper bound performance, it does not reach it. As 
Q increases, the proposed LS solution approaches more closely the sum channel gain achieved by the optimal solution specified in \eqref{eq:pro-formula}. 
Algorithmically, the proposed low-complexity LS method achieves worse sum channel gain than the our proposed Newton-LS method for small $Q$. However, as $Q$ increases, the gap narrows, and at $Q=7$, both methods yield identical performance. Additionally, our proposed Newton-LS method surpasses the Newton-random approach thanks to a better initialization achieved by our proposed LS algorithm.

In Fig.~\ref{fig:num_RIS}, the sum channel gain performance of LS and Newton-LS method is illustrated with different number of BD-RIS elements when $L=K=4$.  As $ N $ increases, the performance of fully connected RIS, Q-stem connected RIS, and the upper bound all improve. Additionally, the performance of the LS method enhances with increasing $ Q $. Specifically, when $ Q = 7 $, the LS method matches the performance of the Newton-LS method. While neither the fully connected RIS nor the Q-stem connected RIS at $ Q = 7 $ reaches the upper bound, but both demonstrate performance comparable to it.

Fig.~\ref{fig:L5K} illustrates the sum channel gain performance of the Newton-LS method as the number of the streams $M$ increases when $L=5$. Given that $K \leq L$, we have $M=K$ in this case. As $M$ grows, the channel gain improves for all BD-RIS architectures. Notably, when $Q=2M-1$,  Q-stem connected RIS matches the performance of  fully connected RIS. 


	\section{Conclusion}
In this paper, we propose a novel Q-stem connected RIS architecture for a downlink BD-RIS-assisted multi-user MISO transmission network, aiming at balancing system performance and RIS circuit complexity. Based on the proposed Q-stem connected RIS, we address its sum channel gain maximization problem by proposing a low-complexity LS algorithm to design its scattering matrix, along with a Newton-based algorithm utilizing our proposed LS solution for initialization. Numerical results demonstrate that our proposed Q-stem connected RIS significantly reduces circuit complexity while maintaining comparable performance to other baseline schemes for some \(Q\) values. Furthermore, the LS method achieves the same performance as fully connected RIS when $Q$ reaches a specific threshold, and the proposed LS-based Newton-based algorithm outperforms other baseline methods significantly. Thus, we conclude that the Q-stem connected RIS and LS algorithm offer substantial potential for practical BD-RIS applications.

	\appendix
When considering a fully connected RIS, i.e., $Q=N-1$, problem \eqref{eq:pro-formula} reduces to
 	\begin{subequations}    
 	\label{eq:fully}
		\begin{align}
			\max_{\bm \Theta} \quad &\|\mathbf{S}_{M}\mathbf{V}_{M}^H\mathbf \Theta \mathbf{P}_{M}\bm \Sigma_M\|_F^2\\
			\operatorname{s.t.} \quad &\bm\Theta\bm\Theta^H=\mathbf I_N, \bm\Theta=\bm\Theta^T.
		\end{align}
	\end{subequations}
Similar to the analysis of problem \eqref{eq:SVD}, the theoretical upper bound of problem \eqref{eq:fully} can be achieved if and only if the following equation set has solution
\begin{subequations}\label{sys of eq}
    \begin{align}
      \label{sys1}  \mathbf{V}_{M}^H\mathbf \Theta \mathbf{P}_{M}&=\bm\Phi,\\
       \label{sys2} \mathbf \Theta\mathbf \Theta^H&=\mathbf I_N,\\
        \mathbf \Theta&=\mathbf \Theta^T.  
    \end{align}
\end{subequations}
The solution of \eqref{sys1} and \eqref{sys2} is given by \begin{equation}\label{eq:theta-opt}
		\bm \Theta^\star=  \mathbf{V}_{M} \bm\Phi\mathbf{P}_{M}^H + \mathbf{V}_{N-M} \mathbf X\mathbf{P}_{N-M}^H,
	\end{equation}
 where $ \mathbf{X}$ is an unitary matrix. Without loss of generality, we set $\phi_m=0$ in the following discussion, as its value does not affect the final objective value. The equation set \eqref{sys of eq} can be further simplified as
\begin{equation}\label{simplified sys}
\bm \Theta=  \mathbf{V}_{M} \mathbf{P}_{M}^H + \mathbf{V}_{N-M} \mathbf{X}\mathbf{P}_{N-M}^H,
        \mathbf \Theta=\mathbf \Theta^T. 
\end{equation}
When considering the special case $ M=1 $, \eqref{simplified sys} has solutions as shown in \cite{nerini2023closed, Santamaria2023}. Next, we prove  by contradiction that when  $ M>1 $,
 \eqref{simplified sys} has no solution.
 
 Assume that there exists at least one  solution of \eqref{simplified sys}, we have 
 \begin{equation}\label{eq:should-sym}
    \begin{split}
         &\mathbf{V}_{M} \mathbf{P}_{M}^H + \mathbf{V}_{N-M} \mathbf{X}\mathbf{P}_{N-M}^H=\mathbf{P}_{M}^*\mathbf{V}_{M}^T + \mathbf{P}_{N-M}^*\mathbf{X}^H\mathbf{V}_{N-M}^T .
    \end{split}
 \end{equation}
By left multiplying $\mathbf{V}_{M}^H$ and right multiplying $\mathbf{V}_{M}^*$ to \eqref{eq:should-sym}, we have
\begin{equation}
    \mathbf{P}_{M}^H \mathbf{V}_{M}^* = \mathbf{V}_{M}^H \mathbf{P}_{M}^*.
\end{equation}
Let $\bm \Lambda \triangleq \mathbf{P}_{M}^H \mathbf{V}_{M}^* \in \mathbb{C}^{M \times M}$, we have $\bm \Lambda=\bm \Lambda^T$. When $M=1$, $\bm \Lambda \in \mathbb{C}$ is a constant scalar, so that $\bm \Lambda=\bm \Lambda^T$ holds obviously. However, when $M>1$, $\bm \Lambda$ is not symmetric with probability $1$ because $ \mathbf{V}_{M} $ and $ \mathbf{P}_{M} $ are unitary matrices from the SVD of $ \mathbf{H}^H $ and $ \mathbf{E} $, respectively. This contradicts our initial assumption.
Therefore, when $ M>1 $, no solution for \eqref{sys of eq} exists. Hence, when $ M>1 $, fully connected RIS cannot achieve the theoretical upper bound performance, completing the proof for the proposition.
	\bibliographystyle{IEEEtran}  
	\bibliography{reference}

\end{document}